\documentclass[11pt,a4paper]{article}

\usepackage{jheppub}
\usepackage{mathtools}
\usepackage{graphicx,amsfonts,color,comment,amsmath,hyperref,float,amssymb}
\usepackage{times}
\usepackage{amssymb}
\usepackage{xspace}
\usepackage{mathrsfs,amssymb}
\usepackage{cancel}
\usepackage[normalem]{ulem}

\newcommand{\eps}{\epsilon}

\newcommand{\be}{\begin{equation}}
\newcommand{\ee}{\end{equation}}
\newcommand{\bea}{\begin{eqnarray}}
\newcommand{\eea}{\end{eqnarray}}

\newcommand{\LMAD}{LMA--Dark\xspace}

\begin{document}

\title{A Plan to Rule out Large Non-Standard Neutrino Interactions After COHERENT Data}

\author[a]{Peter B.~Denton,}

\author[b]{Yasaman Farzan,}

\author[c]{and Ian M.~Shoemaker}

\affiliation[a]{Niels Bohr International Academy, University of Copenhagen, The Niels Bohr Institute,
Blegdamsvej 17, DK-2100, Copenhagen, Denmark}
\affiliation[b]{School of physics, Institute for Research in Fundamental Sciences (IPM)
P.O.Box 19395-5531, Tehran, Iran}
\affiliation[c]{Department of Physics, University of South Dakota, Vermillion, SD 57069, USA}

\emailAdd{peterbd1@gmail.com}
\emailAdd{yasaman@theory.ipm.ac.ir}
\emailAdd{ian.shoemaker@usd.edu}

\date{\today}
\abstract{In the presence of neutrino Non-Standard Interactions (NSI) with matter, the derivation of neutrino parameters from oscillation data must be reconsidered. In particular, along with the standard solution to neutrino oscillation, another solution known as ``LMA--Dark'' is compatible with global oscillation data and requires both $\theta_{12}>\pi/4$ and a certain flavor pattern of NSI with an effective coupling comparable to $G_F$. Contrary to conventional expectations, there is a class of models based on a new $U(1)_X$ gauge symmetry with a gauge boson of mass of few MeV to few 10 MeV that can viably give rise to such large NSI. These models can in principle be tested by Coherent Elastic $\nu$-Nucleus Scattering (CE$\nu$NS) experiments such as COHERENT and the upcoming reactor neutrino experiment, CONUS. We analyze how the recent results from the COHERENT experiment constrain these models and forecast the discovery potential with future measurements from COHERENT and CONUS. We also derive the constraints from COHERENT on lepton flavor violating NSI.}

\preprint{}


\maketitle

\section{Introduction}

When a particle or wave propagates through a medium, due to the collective forward scattering off the particles in the medium, it will feel an effective potential that changes its energy-momentum dispersion relation. In the case of photons, the effect is the well-known refraction phenomenon. Neutrinos propagating in matter undergo a similar effect but given that the interaction is via the weak nuclear force, the speed of neutrinos in matter will remain very close to their speed in vacuum. Nevertheless, the correction to the dispersion relation due to matter effects can impact the pattern of neutrino oscillations which is well-established within the Standard Model (SM) and is a dominant effect for solar neutrinos.

Neutrino oscillation data can also be used to test  the possibility of neutrino interactions with matter fields arising from Beyond the SM (BSM) physics.
Dubbed Non-Standard neutrino Interactions (NSIs), this new physics is typically parameterized by the dimension-6 effective interaction,
\be \mathscr{L}_{NSI} \supset 2 \sqrt{2} G_{F} \epsilon_{\alpha \beta}^{f,V} \left(\bar{\nu}^{\alpha} \gamma^{\mu} \nu_{\beta} \right) \left(\bar{f} \gamma^{\mu} f \right),
\label{LNSI}
\ee
where the parameter $\epsilon_{\alpha \beta}^{f,V}$ determines the strength of the non-standard neutral current interaction between medium fermions $f$ and neutrinos of flavors $\alpha$ and $\beta$ where $\alpha,\beta = (e,\mu,\tau)$. NSI was originally studied in the seminal paper by Wolfenstein on the matter effect~\cite{Wolfenstein:1977ue}, and has since been widely studied in a variety of settings (we refer the reader to the reviews in the literature~\cite{Davidson:2003ha,Ohlsson:2012kf,Farzan:2017xzy}).

As a result of the impact on the matter potential, neutrino oscillation data has provided some of the strongest probes of NSI~\cite{Wolfenstein:1977ue,Roulet:1991sm,Barger:1991ae,Grossman:1995wx,Gonzalez-Garcia:2013usa}. In fact, when neutrino oscillation data is analyzed in the presence of nonzero NSI, in addition to the standard Large Mixing Angle (LMA) solution with $\theta_{12} \simeq 34^{\circ}$ and $\epsilon_{\alpha \beta}^{f} \equiv 0$, another solution, known as \LMAD, appears with $\theta_{12}$ in the ``dark'' octant~\cite{deGouvea:2000pqg} ($45^{\circ} < \theta_{12} < 90^{\circ}$) and large NSI $\epsilon \sim \mathcal{O}(1)$. Distinguishing between the standard LMA solution and this \LMAD~\cite{Miranda:2004nb} regime requires going beyond oscillation data alone.

The most recent probe of NSI comes from the observation of Coherent Elastic $\nu$-Nucleus Scattering (CE$\nu$NS) by the COHERENT experiment~\cite{Akimov:2017ade}. CE$\nu$NS is a process wherein a neutrino scatters coherently off an entire nucleus.
While the cross section is large thanks to the coherent enhancement, $\propto [A-2Z(1-2\sin^2\theta_W)]^2$, it is challenging to detect this process due to the low nuclear recoil energies $\sim$ keV.  The COHERENT collaboration \cite{Akimov:2015nza} reported the first detection of CE$\nu$NS at $6.7$ $\sigma$ \cite{Akimov:2017ade}. The measurement is consistent  with the SM expectations within 1.5 $\sigma$ and therefore offers a new probe of NSI~\cite{Akimov:2017ade,Coloma:2017ncl,Shoemaker:2017lzs,Liao:2017uzy,Farzan:2018gtr}.  Taking the effective interaction of form (\ref{LNSI}), it has been argued that this data is already sufficiently strong to rule out the \LMAD solution~\cite{Coloma:2017ncl}. Notice  however that if the mass of the intermediate state leading to the effective coupling (\ref{LNSI}) is of order of or smaller than the energy-momentum transfer in the scattering experiment, using the effective action formalism will not be viable.

In this paper, we revisit the question of whether or not large NSI can still be accommodated in light of COHERENT data. Our broad conclusion is that it can, though it requires a mediator that is light compared to the momentum transfers probed at COHERENT. We then investigate the possibility of tightening the constraint on LMA-Dark by future CE$\nu$NS results.
The remainder of this paper is organized as follows.  In section \ref{models}, we very briefly describe the class of models that can give rise to \LMAD solution and then in the next section we overview the \LMAD solution phenomenology.
In section \ref{sec:CEvNS}, we discuss the measurement of CE$\nu$NS by COHERENT and use it to constrain the \LMAD solution as well as lepton flavor violating NSI. In section \ref{sec:future},
we  estimate the future sensitivity to the \LMAD solution by both COHERENT and reactor neutrino CE$\nu$NS measurements such as CONUS. Conclusions are summarized in section \ref{CON}.

\section{General characteristics of models leading to large NSI with a light mediator\label{models}}

Similarly to the models developed in \cite{Farzan:2015doa,Farzan:2015hkd,Farzan:2016wym,Farzan:2017xzy}, let us consider an interaction of the following form between neutrinos and quark fields with a new $U(1)_X$ gauge boson, $Z^\prime$  \be
\mathscr{L} \supset \sum_{q \in \{ u,d \}} g_{q} Z'_{\mu} \bar{q} \gamma^{\mu} q + \sum_{\alpha,\beta\in\{e,\mu,\tau\}} (g_{\nu})_{\alpha \beta} Z'_{\mu} \bar{\nu}_\alpha \gamma^{\mu} \nu_\beta.
\label{vectorial} \ee
The coupling of $Z^\prime$ to neutrinos can  originate via (at least) two distinct mechanisms: (1) from gauging an arbitrary (not necessarily flavor universal) linear combination of lepton numbers of different generations \cite{Farzan:2015doa,Farzan:2015hkd}; or, (2) from mixing of $\nu$ with a new electroweak singlet fermion  charged under new $U(1)_X$ with mass of $O$(GeV) \cite{Farzan:2016wym}. The couplings of the quarks to the $Z^\prime$ boson  are $U(1)_X$ gauge couplings. Thus, the flavor structure of $g_q$ is determined by the pattern of the $U(1)_X$ charges assigned to different flavors. For each generation, the $U(1)_X$ charge of the quark with electric charge $2/3$ has to be equal to that of the quark with electric charge $-1/3$ to make the hadronic current coupled to $W_\mu^+$ ({\it i.e.,} $\bar{u} \gamma^\mu(1-\gamma_5) d+ \bar{c} \gamma^\mu(1-\gamma_5) s+\bar{t} \gamma^\mu(1-\gamma_5) b$) invariant under the new $U(1)_X$. As a result from theoretical point of view, we expect
\be \label{geez} g_u=g_d , \ \ \ g_c=g_s \ \ \ {\rm and} \ \ \ g_t=g_b.\ee
Moreover, because of the flavor violation in the mass mixing of quarks ({\it i.e.,} the CKM mixing), any flavor non-universality ($g_u \ne g_c$ and/or $g_u \ne g_t$) can induce
dangerous flavor-changing neutral currents so it will be safer to set $g_u=g_c=g_t$ but this aspect of the model  is not relevant for neutrino oscillation in matter or for CE$\nu$NS experiments in which we are interested in the present paper.

As long as the transferred energy momentum is  small compared to $M_{Z^\prime}$, we can integrate out $Z^\prime$ and arrive at an effective interaction of form Eq.~(\ref{LNSI}) with
\be \epsilon_{\alpha \beta}^q = \frac{(g_\nu)_{\alpha \beta} g_q}{2\sqrt{2} M_{Z^\prime}^2G_F}.\ee
From (\ref{geez}), we conclude 
\be \label{u=d} \epsilon_{\alpha \beta}^u=   \epsilon_{\alpha \beta}^d. \ee
In the literature analyzing the experimental data, it is  however sometimes assumed $\epsilon^u\ne \epsilon^d$, although there is no theoretical justification for this assumption.  

As shown in \cite{Farzan:2015doa,Farzan:2015hkd,Farzan:2016wym,Farzan:2017xzy}, it is possible to reproduce the flavor structure required for the \LMAD solution. Moreover, there are viable mechanism to produce off-diagonal lepton flavor violating as well as lepton flavor conserving $(g_\nu)_{\alpha \beta}$ \cite{Farzan:2015hkd,Farzan:2016wym}.
For neutrino-nucleus scattering experiments (such as COHERENT), the contribution from new interaction to the $\nu$-$N$ scattering amplitude scales as\footnote{Notice that  unlike the case of scalar coupling studied in \cite{Farzan:2018gtr}, with the vectorial interaction that we are considering in Eq.~(\ref{vectorial}), there will be interference between SM contribution and the new physics contribution.}
\be
\label{eq:cx}
\delta \mathcal{M} \propto \begin{cases}
      \frac{g_{\nu}g_{q} }{M_{Z'}^{2}} & \text{if $M_{Z'} \gg q$}, \\
      \frac{g_{\nu}g_{q}}{q^{2}} & \text{if $M_{Z'} \ll q$}.
   \end{cases}
\ee

Independently  of the energy of the neutrino, the non-standard effective potential for neutrinos induced because of the forward scattering of neutrinos off the matter fields in medium  is given by

\be
\label{eq:mat}
(V_{{\rm NSI}})_{\alpha \beta}=
     \sum_{f \in \{ u,d \}} \frac{(g_{\nu})_{\alpha \beta} g_{q}}{M_{Z'}^{2}} N_{f}=2\sqrt{2} G_F \sum_{f\in \{ u, d \}} \epsilon_{\alpha \beta}^f N_f.
\ee

Notice that in forward scattering the energy momentum transfer is zero, $q=0$.  That is why even if the energy of the neutrino beam is larger than the mass of the intermediate state ($M_{Z^\prime}$), for the purpose of calculating the matter effects, we can still use the four-Fermi interaction shown in Eq.~(\ref{LNSI}).
Comparing Eq.~(\ref{eq:cx}) and Eq.~(\ref{eq:mat}), we observe that in the limit $M_{Z^\prime}^2/q^2 \to 0$ and $g_\nu g_q \to 0$ (but fixed $g_\nu g_q/M_{Z^\prime}^2$), the effect on CE$\nu$NS will vanish but still large NSI can be achieved. For a general   matter profile with a given neutron yield $Y_n \equiv N_n/N_p=N_n/N_e$, we can write $(V_{NSI})_{\alpha \beta}=2 \sqrt{2} G_F N_e   \epsilon_{\alpha \beta}$ where\footnote{Throughout the text we distinguish between the Lagrangian level NSI terms (RHS of Eq.~\ref{eps}) from the Hamiltonian level NSI terms (LHS of Eq.~\ref{eps}) by the presence of a quark superscript ($q$, $u$, or $d$) or its absence, respectively.}
\be \label{eps} \epsilon_{\alpha \beta}\equiv (2+Y_n) \epsilon_{\alpha\beta}^{u,V}+ (1+2Y_n) \epsilon_{\alpha\beta}^{d,V}\ .\ee

Before the release of the COHERENT results,
it had been discussed in detail in \cite{Farzan:2015doa,Farzan:2015hkd,Farzan:2016wym,Farzan:2017xzy} that  across the mass window \be 5~{\rm MeV}<M_{Z^\prime}<{\rm few} ~10~{\rm MeV},\label{window}\ee   viable models  respecting all the existing bounds could be built, giving rise to $\epsilon \sim 1$ with
$$\sqrt{g_\nu g_q}\sim 7\times 10^{-5} \sqrt{\epsilon} \frac{M_{Z^\prime}}{10~{\rm MeV}}.$$ The upper limit on the range (\ref{window}) depends on the details of the model. The lower limit of this mass window comes from the  bound on  extra relativistic degrees of freedom from CMB and Big Bang Nucleosynthesis (BBN). As shown in  \cite{Kamada:2015era,Huang:2017egl}, the contribution from $Z^\prime$ to $\delta (N_{\nu})_{eff}$ will violate the bounds if $M_{Z^\prime}<5$~MeV  and $g_\nu >10^{-9} (M_{Z^\prime}/{\rm MeV})$.  This constraint is obtained by studying the thermalization and decay of the $Z'$. Even if the mass of $Z'$ is large enough to make $Z^\prime$  non-relativistic at the neutrino decoupling era, its subsequent decay into a neutrino pair can effectively heat the neutrino bath.

In the parameter range of our interest, the $Z^\prime$ boson can be produced inside the supernova core and decay back to a neutrino/antineutrino pair within the core. This production cannot provide a new cooling mechanism for the star but by providing a new neutrino scattering channel it can affect the duration of the neutrino emission. Any direct information from CE$\nu$NS on the $Z^\prime$ coupling to $\nu$ would be an invaluable input for studies of supernova and for predicting the neutrino emission duration.

We also note that both oscillation experiments and scattering experiments are only sensitive to the product $g_\nu g_q$.
It may be possible to constrain the $g_\nu$ term directly (and therefore constrain $g_q$ through the combination) through Non-Standard neutrino Self-Interactions (NSSI) from the measurement of the neutrino spectra from a galactic supernova \cite{Das:2017iuj}. Moreover, rare meson decays can constrain $g_\nu$ \cite{Bakhti:2017jhm}.

In this work we restrict ourselves to vector NSI with quarks only and most of the time drop the superscript $V$ from $\epsilon^V$.
Axial-vector NSI are fairly well constrained at the $\eps^A\sim0.1$ level from SNO neutral current measurements \cite{Miranda:2004nb}.

\section{\LMAD}
\label{sec:lmad}
In this section we review the theoretical derivation of the \LMAD solution and then describe the latest constraints from oscillation experiments determined in a global fit by Ref.~\cite{Coloma:2017egw}.

\subsection{\LMAD theory review}

The CPT invariance implies the invariance of  the neutrino Hamiltonian  under $H\to-H^*$, leading to the Generalized Mass Ordering Degeneracy (GMOD) \cite{Coloma:2016gei}.
In vacuum this leads to the \LMAD solution wherein $\theta_{12}>45^\circ$, degenerate with the standard LMA solution \cite{deGouvea:2000pqg}.
In matter the degeneracy is broken, but can be restored with new physics in the form of NSI of the same magnitude as the weak scale, $\eps=\mathcal O(1)$ \cite{Miranda:2004nb}.
In particular, if $\eps_{ee}=-2$, the $ee$ term of the matter potential changes sign maintaining the degeneracy.
Furthermore, adding any term proportional to the identity matrix to the $3\times 3$ Hamiltonian of neutrinos does not affect neutrino oscillations.
Thus, as far as neutrino oscillations are concerned, the SM is equivalent to $(\eps_{ee},\eps_{\mu\mu},\eps_{\tau\tau})=(-2,0,0)$ as well as $(0,2,2)$ or any expression of the form
\begin{equation}
(\eps_{ee},\eps_{\mu\mu},\eps_{\tau\tau})=(x-2,x,x)\,, \label{LMA-pat}
\end{equation}
for arbitrary real $x$.
Since the neutrino beam at the COHERENT experiment is composed of both $\nu_\mu$ and $\nu_e$ fluxes, its sensitivity to $x$ is almost flat but the reactor CE$\nu$NS experiments, having only $\bar{\nu}_e$ beam, will  lose sensitivity at $x=2$.

By looking at oscillations in different matter densities with different neutron to proton ratios, the GMOD can be broken again, except for the case where the neutron contribution is zero.
From Eq.~(\ref{eps}), we observe that vanishing neutron contribution requires $\epsilon_{\alpha \beta}^{u,V}+2 \epsilon_{\alpha \beta}^{d,V}=0$. Thus, no oscillation experiment can distinguish between standard LMA solution and the LMA-Dark solution with $\epsilon_{ee}^{d,V}=-(x-2)/3$, $\epsilon_{ee}^{u,V}=2(x-2)/3$, $\epsilon_{\mu\mu}^{d,V}=\epsilon_{\tau\tau}^{d,V}=-x/3$ and
$\epsilon_{\mu\mu}^{u,V}=\epsilon_{\tau\tau}^{u,V}=2x/3$.
Notice however that within the models described in section \ref{models}, we expect $\epsilon^u=\epsilon^d$.

Scattering experiments are required to break these degeneracies.
While oscillations constrain NSI for any mediator mass, scattering experiments can only constrain NSI when the transfer energy is less than the mediator mass $q\lesssim M_{Z'}$.
Scattering experiments and oscillation experiments are therefore complementary: while the oscillation experiments can constrain NSI for any mediator mass, but are insensitive to the $x$ parameter of Eq.~(\ref{LMA-pat}) and the GMOD, the scattering experiments can break these degeneracies, but are only sensitive to certain mediator mass ranges.

\subsection{Oscillation constraints on \LMAD}

From a global fit to neutrino oscillation data, Ref.~\cite{Coloma:2017egw} obtains the 90\% C.L.~limits shown in Table \ref{tab:oscillation NSI limits}.
From Eq.~(\ref{eps}) along with the one-at-a-time values in Table \ref{tab:oscillation NSI limits} we can observe that the \LMAD solution found in oscillations dominantly comes from data with $Y_n<1$ implying that the solar data dominates the contribution to the \LMAD solution, as expected. Unless stated otherwise, from hereon whenever we discuss \LMAD solution we set $x=0$ ({\it i.e.,} $\epsilon_{ee}=-2$ and $\epsilon_{\mu\mu }=\epsilon_{\tau\tau }=0$).

\begin{table}
\centering
\caption{Limits at 90\% C.L.~on NSI from a global fit to neutrino oscillation data while marginalizing over all other standard and NSI parameters taken from \cite{Coloma:2017egw}.
The marginalizations are performed leaving NSI with one quark ($q=u,d$) at a time free.
The $\epsilon\sim-1$ solutions corresponds to the LMA--Dark solution with $\theta_{12}>45^\circ$.}
\label{tab:oscillation NSI limits}
\begin{tabular}{c|c|c}
&$\eps_{ee}^{q,V}-\eps_{\mu\mu}^{q,V}$&$\eps_{\mu\mu}^{q,V}-\eps_{\tau\tau}^{q,V}$\\\hline
$q=u$&$[-1.19,-0.81]\oplus[0.00,0.51]$&$[-0.03,0.03]$\\
$q=d$&$[-1.17,-1.03]\oplus[0.02,0.51]$&$[-0.01,0.03]$
\end{tabular}
\end{table}

\begin{figure}
\centering
\includegraphics[width=3in]{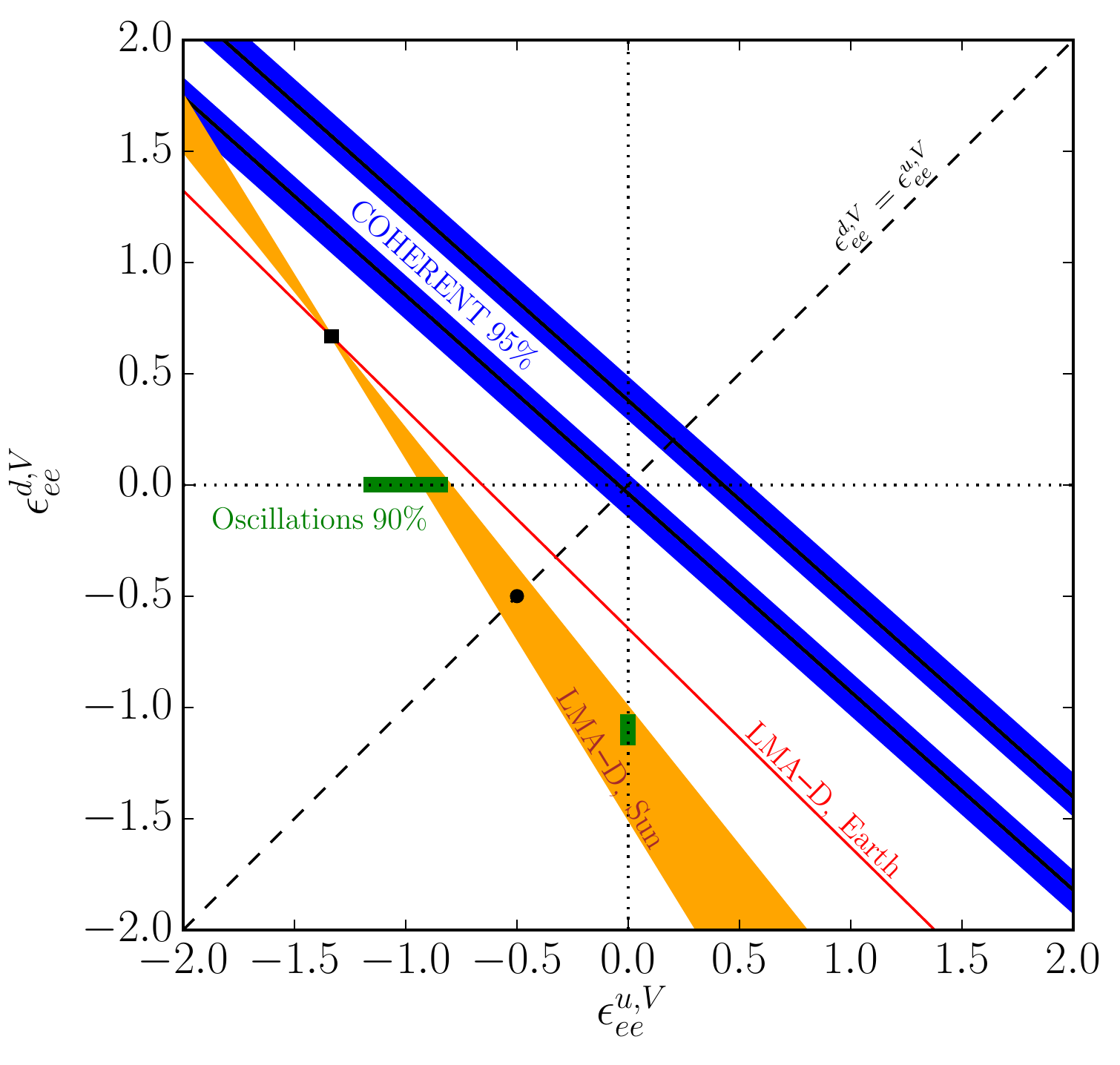}
\caption{
The constraints on NSI in the $\epsilon_{ee}^{d,V}$ -- $\epsilon_{ee}^{u,V}$ plane, setting all other NSI terms  to  zero.
The red line and the orange region show $\epsilon_{ee}=-2$ respectively for $Y_n=1.05$ (the value in the Earth) and for $Y_n\in[1/6,1/2]$ (the values in the Sun).
The intersection of these two lines shown by the square point is the point at which oscillations are exactly degenerate.
The circle is the point on the $\epsilon_{ee}^{d,V}=\epsilon_{ee}^{u,V}$ line we take for our canonical LMA--Dark value.
The best fits values from COHERENT at $\chi^2=2.9$ are the black lines with the 95\% C.L.~(2 d.o.f.) region shown in blue assuming $x=0$ and large $M_{Z'}$.
The green bands represent the one at a time LMA--Dark limits from the oscillation data in Table \ref{tab:oscillation NSI limits} from \cite{Coloma:2017egw} confirming that solar data dominates the \LMAD constraint.}
\label{fig:ee_u_d}
\end{figure}

As mentioned above, we focus on models with $\eps_{\alpha\beta}^{u,V}=\eps_{\alpha\beta}^{d,V}$.
The \LMAD solution ($\epsilon_{ee}=-2$) then results in
\begin{equation}
(1+Y_n)\eps_{ee}^{q,V}=-\frac23\,. \label{y-n}
\end{equation}
Since $Y_n$ varies in the range $[1/6,1.05]$ which is the experimentally probed range, we choose $Y_n=1/3$ which is in the middle of solar range ($Y_n\in[1/6,1/2]$) because  as shown in Fig.~\ref{fig:ee_u_d}, the solar data provides  the main constraint on \LMAD.
This gives our canonical definition of \LMAD of $\eps_{ee}^{u,V}=\eps_{ee}^{d,V}=-1/2$, although we also consider varying $x$ as defined in Eq.~(\ref{LMA-pat}).
While the red line (marked with LMA-D, Earth) and orange region (marked with LMA-D, Sun) are the solutions to $\epsilon_{ee}=-2$ for relevant values of $Y_n$, the green
regions are observational limits, derived from data. Notice that the uncertainties in the current atmospheric and long baseline neutrino data are too large to allow sensitivity to matter effects.
In fact, the observational constraint on the \LMAD solution comes mainly from solar data. This is confirmed by the overlap of the green regions (corresponding to the one at a time global fit limits from Table \ref{tab:oscillation NSI limits}) with the orange region as well as the absence of any overlap with the red line.

\section{Coherent Elastic \texorpdfstring{$\nu$}{nu}-Nucleus Scattering}
\label{sec:CEvNS}
\subsection{COHERENT constraints on the LMA--Dark solution}
As was pointed out in \cite{Scholberg:2005qs,Coloma:2017egw}, a Coherent Elastic $\nu$-Nucleus Scattering (CE$\nu$NS pronounced ``sevens") experiment such as COHERENT could be used to constrain NSI for light mediators with masses $\mathcal O(10)$ MeV.
Above $\sim1$ GeV additional Deep-Inelastic Scattering (DIS) constraints from CHARM \cite{Dorenbosch:1986tb} and NuTeV \cite{Zeller:2001hh} apply, with the NuTeV constraints being particularly strong \cite{Coloma:2017egw}.
The recent COHERENT data has been used to constrain NSI for $M_{Z'}>\mathcal O(10)$ MeV \cite{Akimov:2017ade,Liao:2017uzy,Coloma:2017ncl}.
We expand upon those analyses here with a focus on the \LMAD solution.

CE$\nu$NS is a  process wherein a neutrino scatters elastically off an entire nucleus.
Thus, the scattering cross section will be given by the square of the sum of the scattering amplitudes off each nucleon in the nucleus and scales with square of atomic number.
Within the standard model, the cross section is enhanced by $[A-2Z(1-2\sin^2 \theta_W)]^2$ and is relatively large. However, it is difficult to detect CE$\nu$NS due to low nuclear recoil energies $\sim$ keV.
Recently the COHERENT collaboration \cite{Akimov:2015nza} reported the first detection of CE$\nu$NS at $6.7$ $\sigma$ \cite{Akimov:2017ade}.
COHERENT uses neutrinos from pion decay at rest (DAR) coming from the Spallation Neutron Source (SNS) at Oak Ridge National Laboratory detected in a low threshold CsI detector.

We calculate the CE$\nu$NS event rates as a function of the NSI parameters as described in \cite{Coloma:2017egw} using form factors from \cite{Klein:1999gv} and a detection threshold of 7 keV \cite{Collar:2014lya}.
We assume  the background to be $20\%$ of the signal and a systematic uncertainty in the total flux of $20\%$ consistent with the uncertainties reported by COHERENT.
We marginalize the $\chi^2$ over the normalization uncertainty using the pull method \cite{Fogli:2002pt}.

The SNS beam is pulsed which means that the $\nu_\mu$'s from the prompt $\pi^+$ decay can be distinguished from the delayed $\nu_e$'s and $\bar{\nu}_\mu$'s from the $\mu^+$ decay coming from the initial $\pi^+$ decay.
We make use of two separate timing bins contributing to the $\chi^2$ as first described in \cite{Coloma:2017egw}: the prompt component and delayed components.
The numbers of prompt and delayed events, as a function of each flavor are
\begin{gather}
\begin{aligned}
N_p&=N_{\nu_\mu}+P_c(N_{\nu_e}+N_{\bar{\nu}_\mu})\,,\\
N_d&=(1-P_c)(N_{\nu_e}+N_{\bar{\nu}_\mu})\,,
\end{aligned}
\label{eq:Np Nd}
\end{gather}
where the contamination from early muon decay given by
\begin{equation}
P_c=\frac1{p_w}\int_0^{p_w}dt[1-e^{-(b_w-t)/\Gamma\tau}]=0.246\,,
\label{eq:Pc}
\end{equation}
in which $p_w=0.695$ $\mu$s is the pulse width and $b_w=1$ $\mu$s is the bin width from the data presented by COHERENT.
Note that our results are fairly insensitive to the value of $P_c$; as long as the prompt and delayed events can be largely separated, we get the full benefit of discriminating between the flavors.
The contamination due to other backgrounds are suppressed by at least two orders of magnitude and are safely ignored here.

The per-flavor event rates are then given by
\begin{equation}
N_\alpha=N_t\Delta t\frac{G_F^2}{2\pi}M_t\int_{E_{r,{\rm tr}}}dE_r\int dE_\nu\phi_\alpha(E_\nu)\frac{Q_{w\alpha}^2(\sqrt{2M_tE_r})}4F^2(2M_tE_r)\left(2-\frac{M_tE_r}{E_\nu^2}\right)\,,
\label{eq:N}
\end{equation}
where $M_t$ is the mass of the target nuclei, $N_t$ is the number of target nuclei in the detector, and $E_{r,{\rm tr}}$ is the threshold recoil energy.
The electroweak charge is
\begin{equation}
\frac{Q_{w\alpha}^2(q)}4=\left[Zg_p^V+Ng_n^V+3(Z+N)\eps_{\alpha\alpha}^{q,V}(q)\right]^2
+9(Z+N)^2\sum_{\beta\neq\alpha}\left[\eps_{\alpha\beta}^{q,V}(q)\right]^2\,,
\label{eq:weak charge}
\end{equation}
and the normalized per-flavor fluxes from $\pi$DAR are to an excellent approximation given by kinematics as
\begin{gather}
\begin{aligned}
f_{\nu_\mu}&=\delta\left(E_\nu-\frac{m_\pi^2-m_\mu^2}{2m_\pi}\right)\,,\\
f_{\bar\nu_\mu}&=\frac{64}{m_\mu}\left[\left(\frac{E_\nu}{m_\mu}\right)^2\left(\frac34-\frac{E_\nu}{m_\mu}\right)\right]\,,\\
f_{\nu_e}&=\frac{192}{m_\mu}\left[\left(\frac{E_\nu}{m_\mu}\right)^2\left(\frac12-\frac{E_\nu}{m_\mu}\right)\right]\,,
\end{aligned}
\label{eq:fluxes}
\end{gather}
where $E_\nu\in[0,m_\mu/2]$.
In general we fix all off-diagonal NSI terms to be zero unless otherwise specified.
Note that there is a degeneracy in the weak charge between the SM and NSI which occurs at
\begin{equation}
\eps_{\alpha\alpha}^{q,V}(q)=-\frac{2(Zg_p^V+Ng_n^V)}{3(Z+N)}\,.
\label{eq:degeneracy}
\end{equation}
For COHERENT, this corresponds to $\eps_{\alpha\alpha}^{q,V}=0.18$ in the heavy mediator limit for $g_n^V=-\frac12$ and $g_p^V=\frac12-2\sin^2\theta_W\approx0.055$.

The current COHERENT constraints in the $\eps_{ee}$ sector are shown in Fig.~\ref{fig:ee_u_d} for heavy mediator at $x=0$.
Note that these results are stronger than those previously presented \cite{Akimov:2017ade} due to the additional timing information to separate electron and muon neutrinos.
While the SM ($\eps=0$) is included within the blue bands, it is disfavored.
A good fit with $\chi^2=0$ can be obtained by varying more than just the $\eps_{ee}^{q,V}$ terms.

For COHERENT to be sensitive to the details of the $Z'$, there must be nonzero momentum transfer. This leads us to define the generalized NSI coefficient,
\begin{equation} \label{q-dependent}
\eps_{\alpha\beta}^{f,V}(q)\equiv \frac{(g_\nu)_{\alpha\beta}g_f}{2\sqrt2G_F(q^2+M_{Z'}^2)}= \eps_{\alpha\beta}^{f,V}(0)\frac{M_{Z^\prime}^2}{q^2+M_{Z^\prime}^2}\,,
\end{equation}
which is related to the $\epsilon$'s relevant to oscillation physics by taking the $q=0$ limit, $\eps_{\alpha\beta}^{f,V}\equiv\eps_{\alpha\beta}^{f,V}(q=0)$.

For $M_{Z^\prime} \sim q$, in principle by studying the energy dependence of the scattering cross section, the values of both
$M_{Z^\prime}$ and the product $(g_\nu)_{\alpha \beta} g_f$ can be extracted. Taking a flavor universal coupling to neutrinos and using the released COHERENT data, Ref.~\cite{Liao:2017uzy}
constrains $\sqrt{g_\nu g_q}$ for $M_{Z^\prime} \sim $ few~10~MeV. In principle, by using the timing information to discriminate between flavors a similar analysis of energy spectrum can be carried out for arbitrary flavor structure of NSI including the \LMAD flavor pattern in Eq.~(\ref{LMA-pat}). Although the COHERENT collaboration has released the information both on time (count per arrival time bin) and on energy (count per number of photoelectrons), it has not unfortunately released information on simultaneous dependence on both (count per time per number of the photoelectrons). In the absence of this information, we have resorted to using only the timing (or equivalently only flavor information) to derive bounds on $M_{Z^\prime}$. In the event that COHERENT releases the energy spectrum in both timing bins, we expect that even stronger constraints could be placed by combining timing and energy information.

Taking the \LMAD solution ({\it i.e.,} flavor pattern in Eq.~(\ref{LMA-pat})) with various values of $x$ and $Y_n=1/3$ (the average neutron yield in the Sun), we have computed $\epsilon_{\alpha \beta}^{f,V}(q)$ in terms of $M_{Z^\prime}$  and calculated $\chi^2$ defined as
\begin{equation}
\chi^2=\min_{x,\xi}\sum_{k=p,d}\left[\frac{(1+\xi)N_{k,{\rm NSI}}(x)-N_{k,{\rm obs}}}{\sqrt{N_{k,{\rm obs}}+0.2N_{k,obs}}}\right]^2+\left(\frac\xi{\sigma_{\rm sys}}\right)^2\,,
\end{equation}
where $k\in{p,d}$ is the set of prompt and delayed signals, the 0.2 represents the $20\%$ background rate, and we take $\sigma_{\rm sys}=0.2$ for the systematic normalization uncertainty.
The event rates are defined in Eqs.~(\ref{eq:Np Nd}--\ref{eq:fluxes}).

\begin{figure}
\centering
\includegraphics[width=3in]{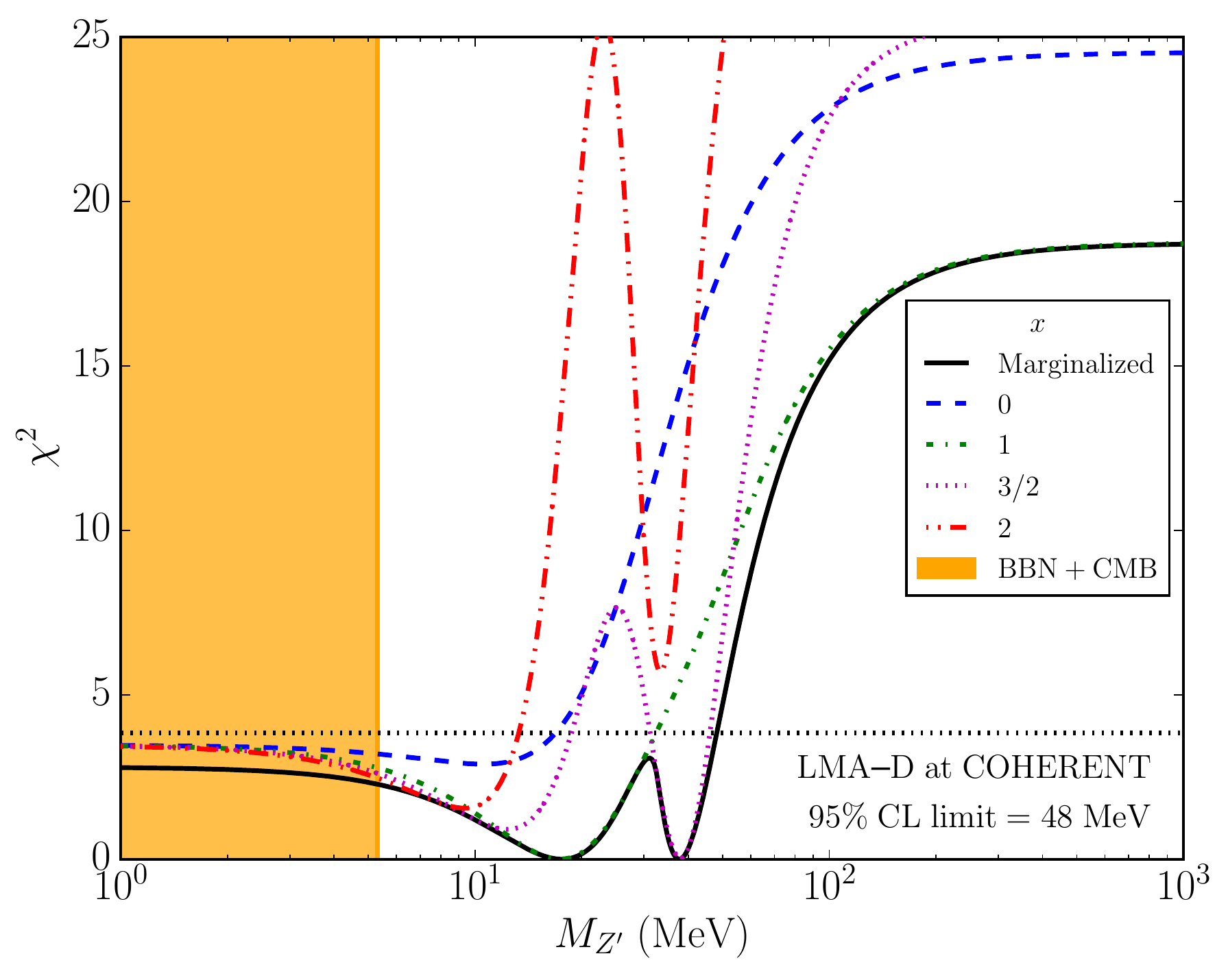}
\caption{The $\chi^2$ of the COHERENT data, using timing information, at $\epsilon_{ee}^{u,V}=\epsilon_{ee}^{d,V}=\frac x4-\frac12$ as a function of $M_{Z'}$.
The black curve includes a marginalization over $x$, while the other curves show the constraint for various values of $x$ from Eq.~(\ref{LMA-pat}).
The orange shaded area in the left is excluded by the bound on extra relativistic degrees of freedom from BBN and the CMB \cite{Kamada:2015era}. The horizontal line shows the 95\% C.L.~limit for 1 d.o.f.
The data rules out LMA--Dark for $M_{Z'}>48$ MeV at 95\% C.L.~(1 d.o.f.).
Note that the $\chi^2$ remains non-zero as $M_{Z'}\to0$ because the COHERENT measurement is slightly off the SM prediction at low significance.}
\label{fig:LMAD_Exact_MZ}
\end{figure}

The $\chi^2$ for the \LMAD solution as a function of mediator mass is shown in Fig.~\ref{fig:LMAD_Exact_MZ}.  Notice that for fixed $(\epsilon_{ee},\epsilon_{\mu\mu},\epsilon_{\tau\tau})$,  $M_{Z^\prime} \to 0$ corresponds to the SM with $g_\nu g_q \to 0$. Had the best fit of the COHERENT data corresponded to the SM prediction, the $\chi^2$ would have approached zero as $M_{Z^\prime} \to 0$. The SM prediction however has a small (1.5 $\sigma$ C.L.) deviation from the results of COHERENT and this justified convergence to a nonzero value of $\chi^2$ at $M_{Z^\prime} \to 0$. From Fig.~\ref{fig:LMAD_Exact_MZ}, we observe that for all values of $x$ considered, there are dips which means the corresponding NSI can provide better fit to data than the SM (the limit $\epsilon(q^2) \to 0$). For $x=3/2$ and $x=1$, the $\chi^2$ can even vanish at $M_{Z'}=38$ MeV and $18$ MeV respectively. The solid black curve is the result of marginalizing over $x$.  As seen from this figure, COHERENT constrains NSI \LMAD for mediator masses $M_{Z'}>48$ MeV at 95\% C.L.~after marginalizing over $x$.
This constraint is dominated by $x\approx3/2$ or $(\eps_{ee},\eps_{\mu\mu},\eps_{\tau\tau})=(-1/2,3/2,3/2)$.
If we fix $x=0$, the constraint improves to 17 MeV.
The multiple dip structure is a result of the fact that the event rate scales roughly like $[g_{\rm SM}+\eps(q)]^2$ where $\eps(q)$ is a function of both  $M_{Z'}$ and $x$  (through $\eps(0)$); see  Eqs.~(\ref{LMA-pat},\ref{q-dependent}).

\subsection{Additional COHERENT constraints}
\begin{figure}
\centering
\includegraphics[width=3.5in]{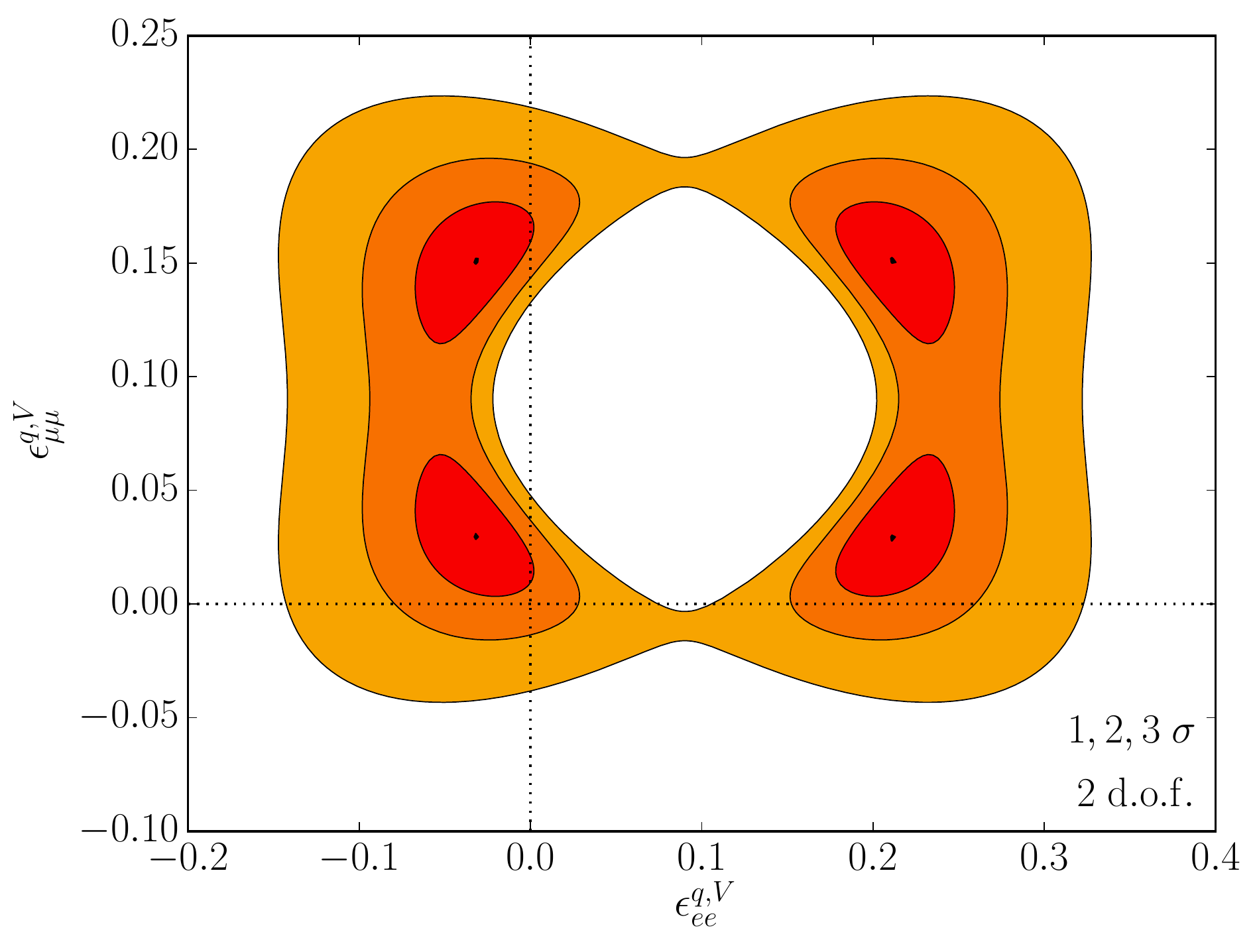}
\caption{The 1, 2, and 3 $\sigma$ constraints (2 d.o.f.) for the two diagonal NSI terms from COHERENT's measurement using timing information where all other NSI terms are set to zero.
We have assumed that $\epsilon^u=\epsilon^d$ as required by underlying models for NSI and have taken the mediator to be heavy.
At the best fit points the $\chi^2=0$.}
\label{fig:ee_mm}
\end{figure}

Beyond constraining large NSI in the form of \LMAD, COHERENT can also constrain the NSI parameters directly.
Maintaining $\eps^u=\eps^d$, COHERENT can constrain the $ee$ and $\mu\mu$ elements as shown in Fig.~\ref{fig:ee_mm}.
COHERENT has no sensitivity to the $\tau$ sector, but constraints can be inferred by including oscillation information (see Table \ref{tab:oscillation NSI limits}) which constrains $|\eps_{\mu\mu}^{q,V}-\eps_{\tau\tau}^{q,V}|\lesssim0.03$, so the bounds on $\eps_{\mu\mu}^{q,V}$ are essentially the same as those on $\eps_{\tau\tau}^{q,V}$.
Note that there are four points where the $\chi^2=0$.
These are related to the degeneracy mentioned in Eq.~(\ref{eq:degeneracy}), but are not quite at exactly 0.18 since COHERENT did not measure the SM.
Had COHERENT measured the SM, all four would be at $\eps_{\alpha\alpha}^{q,V}=0.18$.

\begin{figure}
\centering
\includegraphics[width=4.5in]{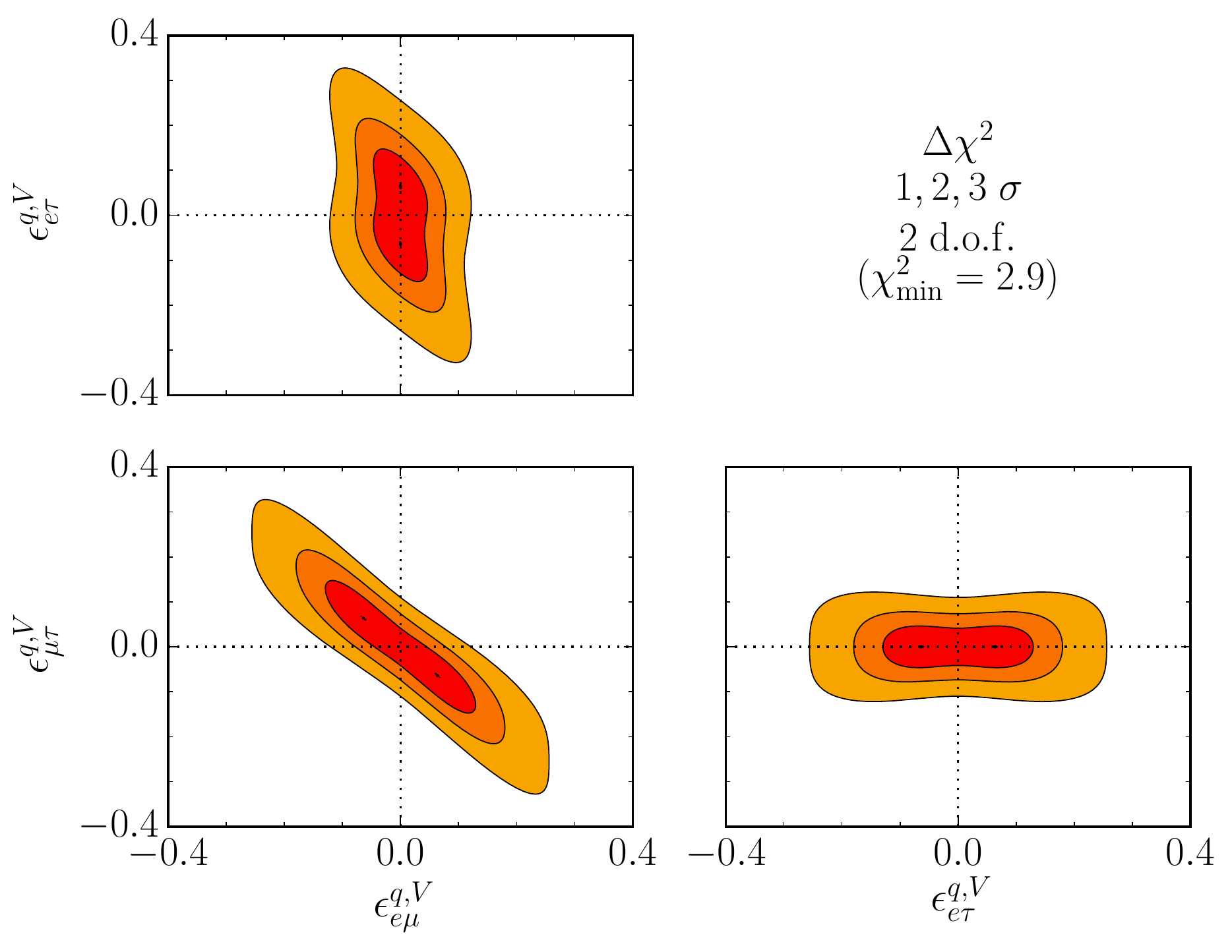}
\caption{The 1, 2, and 3 $\sigma$ $\Delta\chi^2$ constraints (2 d.o.f.) for off-diagonal NSI from COHERENT's measurement using timing information where all other NSI terms are set to zero.
We have assumed that $\epsilon^u=\epsilon^d$ as required by underlying models for NSI and have taken the mediator to be heavy.
The minimum $\chi^2$ is $2.9$.}
\label{fig:Triangle}
\end{figure}

The COHERENT experiment also constrains the off-diagonal NSI terms $\eps^{q,V}_{e\tau}$, $\eps^{q,V}_{\mu\tau}$ and $\eps^{q,V}_{e\mu}$ as shown in Fig.~\ref{fig:Triangle}.
One at a time constraints are listed in Table \ref{tab:COHERENT NSI limits}.
COHERENT is able to constrain all the NSI parameters except for the $\tau\tau$ term.
Constraining the $\tau\tau$ element is possible by  combining the bound on the $\mu \mu$ component from the COHERENT with the $|\epsilon_{\mu\mu}-\epsilon_{\tau\tau}|\lesssim0.03$ constraint from oscillations listed in Table \ref{tab:oscillation NSI limits}.

\begin{table}
\centering
\caption{One at a time constraints at 90\% C.L.~from COHERENT alone for NSI with a heavy mediator assuming that $\epsilon^u=\epsilon^d$.}
\label{tab:COHERENT NSI limits}
\begin{tabular}{c|c}
&$\eps_{\alpha\beta}^{q,V}$\\\hline
$ee$&$[-0.073,0.023]\oplus[0.16,0.25]$\\
$\mu\mu$&$[-0.0070,0.033]\oplus[0.15,0.19]$\\
$e\mu$&$[-0.055,0.055]$\\
$e\tau$&$[-0.014,0.014]$\\
$\mu\tau$&$[-0.051,0.051]$
\end{tabular}
\end{table}

\section{Future expectations}
\label{sec:future}
\subsection{\texorpdfstring{$\pi$}{pi}-Decay At Rest: COHERENT}
\begin{figure}[t!]
\centering
\includegraphics[width=3in]{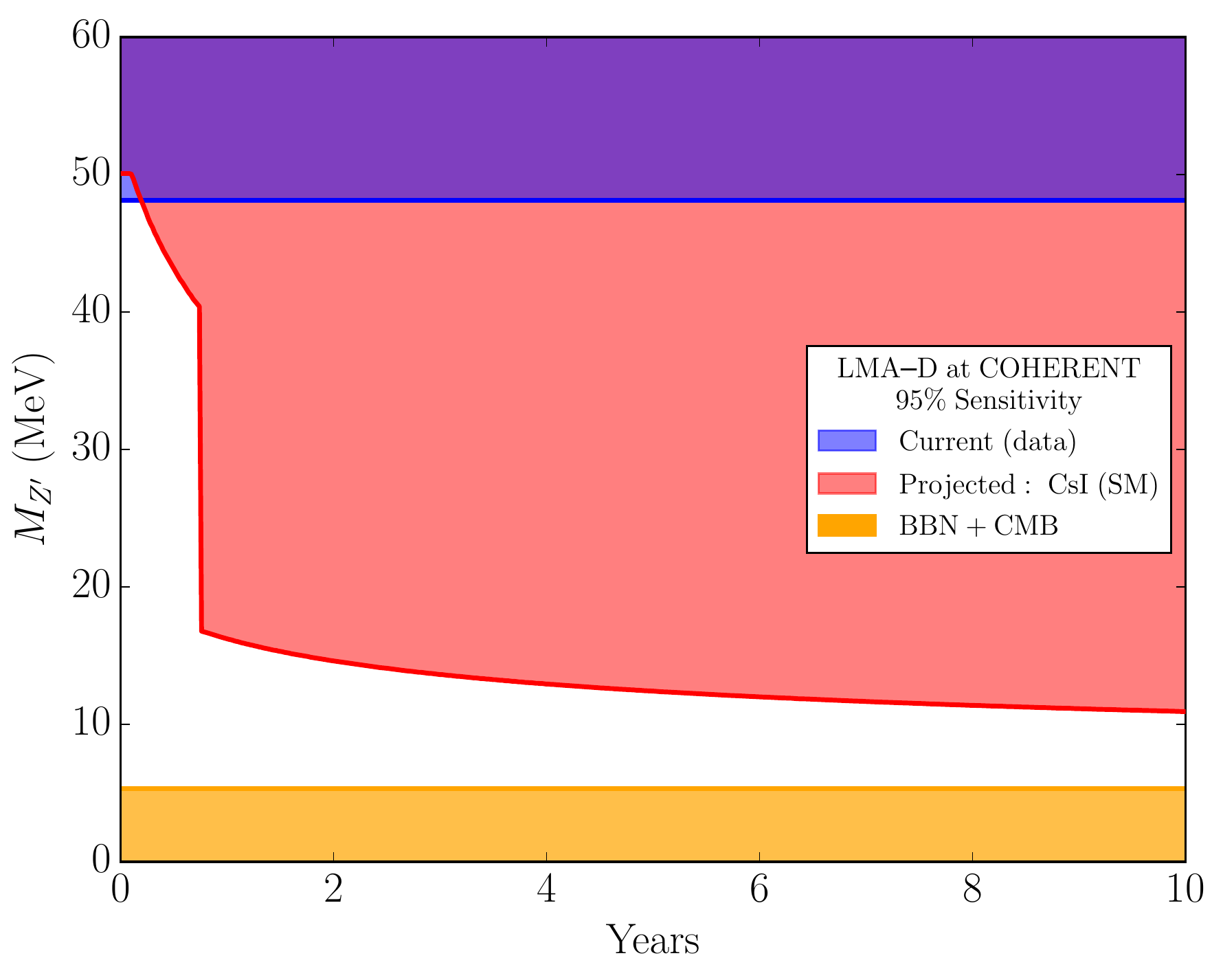}
\caption{Future sensitivity at 95\% C.L.~of COHERENT to the exactly degenerate \LMAD solution for different NSI mass scales $M_{Z'}$ including the marginalization over $x$.
The horizontal axis shows the real time, and we assume 50\% uptime.
The blue region is the current exclusion limit as shown in Fig.~\ref{fig:LMAD_Exact_MZ}.
The red region is the predicted future exclusion range assuming true value of $\epsilon=0$ which becomes saturated at $\sim10$ MeV.
The sharp drop occurs when the higher mass minimum seen in Fig.~\ref{fig:LMAD_Exact_MZ} passes the threshold.
The orange region is the exclusion limit coming from BBN and CMB constraints \cite{Kamada:2015era}.
Future measurements from reactor experiments like CONUS will reach the $\sim1$ MeV level and this figure will be completely covered.}
\label{fig:MZ_Sensitivity}
\end{figure}

Assuming COHERENT's CsI detector continues at its current rate\footnote{As COHERENT continues taking data, they will also be adding additional detector materials \cite{Akimov:2018ghi}.
Materials with different neutron to proton ratios (down to up quark ratios) will improve their precision, particularly for $\eps_{ee}$, although at the current statistical and systematics level, the improvement will be marginal and largely statistical.} and collects data $\sim$ half the time, the expected future sensitivity of COHERENT to $M_{Z'}$ for the \LMAD solution is shown in Fig.~\ref{fig:MZ_Sensitivity} which also includes a marginalization over $x$.
Two features are of note.
The first is the sharp improvement in the sensitivity.
This is due to the non-trivial shape of the exclusion plot shown in Fig.~\ref{fig:LMAD_Exact_MZ}.
When the dip in the $\chi^2$ increases past the threshold, the sensitivity suddenly improves considerably.
The other feature is that the current projected limit is slightly worse than the actual current limit.
This is because for the sensitivity we have assumed that COHERENT will exactly measure the SM: $\eps=0$, while their current measurements are slightly higher than the SM leading to slightly different limits.

\subsection{Reactor: CONUS}
\label{ssec:CONUS}
Reactor neutrinos will also help to constrain NSI \cite{Barranco:2005yy,Dent:2017mpr,Farzan:2018gtr} and numerous such experiments are in various stages of progress from running to proposed including TEXONO, NOSTOS, CONUS, GEMMA, CONNIE, MINER, and others \cite{Wong:2005vg,Aune:2005is,Brudanin:2014iya,Lindner:2016wff,Aguilar-Arevalo:2016qen,Agnolet:2016zir,Billard:2016giu}.
One such experiment is the COhernt NeUtrino Scattering experiment (CONUS), a proposed experiment to measure CE$\nu$NS from reactor neutrinos with a Germanium detector and an ultra-low threshold of $\sim0.1$ keV. They anticipate $\sim10^5$ events assuming standard physics over five years \cite{Lindner:2016wff}.

\begin{figure}
\centering
\includegraphics[width=3in]{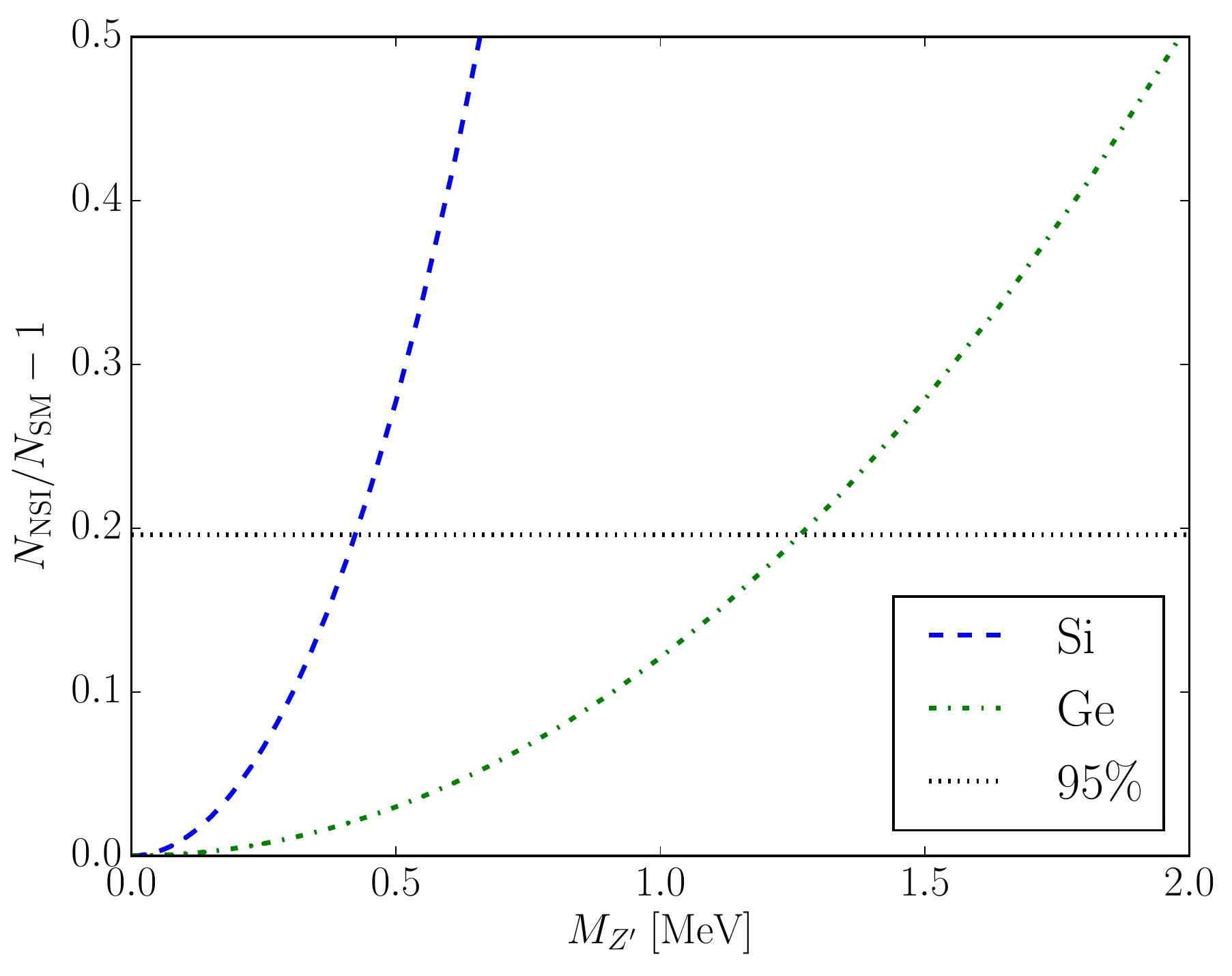}
\caption{The deviation of the prediction of the LMA--Dark solution with mediator of mass $M_{Z^\prime}$ from that of SM in a reactor neutrino setup such as CONUS.
The Si and Ge detectors are taken to have recoil energy thresholds of $0.1$ keV and $0.6$ keV respectively.
The horizontal dotted line shows the 95\% C.L.~bound, assuming $10^5$ events in case of the SM and constrains $M_{Z'}<0.45$ MeV ($1.3$ MeV) with Si (Ge) detectors respectively.
Bounds from BBN and the CMB at $M_{Z'}>5.3$ MeV \cite{Kamada:2015era} already rule out the entire $M_{Z'}$ range in this figure.}
\label{fig:reactor MZ sensitivity}
\end{figure}

We simulate the expected signal for the SM and for \LMAD with different mediator masses.
We take the $^{235}$U flux from \cite{Huber:2004xh} and form factors from \cite{Klein:1999gv}, although the suppression from form factors are negligible since $F(q^2)\sim1$ for relevant energies.
We conservatively estimate the systematic uncertainty from various reactor neutrino uncertainties and detector uncertainties to be $10\%$ to account for nuclear uncertainties, the reactor anomaly \cite{Mention:2011rk}, and the 5 MeV bump \cite{RENO:2015ksa}, and we consider a count only analysis.\footnote{A shape analysis is possible as well since NSI does modify the spectrum, but this is not included in this work.}
With $10^5$ events the result is completely dominated by systematics.
Assuming these detectors measure the SM ($\epsilon=0$), their ability to constrain the \LMAD with  $x=0$ ({\it i.e.,} $(\epsilon_{ee},\epsilon_{\mu \mu},\epsilon_{\tau\tau})=(-2,0,0)$)  is shown in Fig.~\ref{fig:reactor MZ sensitivity}.
The Si and Ge detectors respectively impose 
$M_{Z'}<0.45$ MeV and $M_{Z'}<1.3$ MeV at 95\% C.L.
The difference is dominated by the choice of detector nuclear recoil thresholds, $0.1$ keV and $0.6$ keV for Si and Ge respectively. Recall that at $x=3/2$, the constraint by COHERENT was the weakest providing an upper bound  $M_{Z'}<48$ MeV.
At  $x=3/2$,  CONUS with Si and Ge detectors can  constrain the LMA-Dark solution with light mediator respectively to $M_{Z'}<0.9$ MeV and $2.6$ MeV, both of which are well below the constraint from BBN and the CMB covering the gap.
In addition, for comparison, in the event that the flux uncertainties can be reduced to optimistic levels of $1\%$,  the constraints improve to $0.15$ and $0.45$ MeV for Si and Ge respectively.
We note that these results are quite general and apply to a wide range of possible detectors, limited  mainly by the flux uncertainties.

\bigskip

\begin{figure}
\centering
\includegraphics[width=0.49\textwidth]{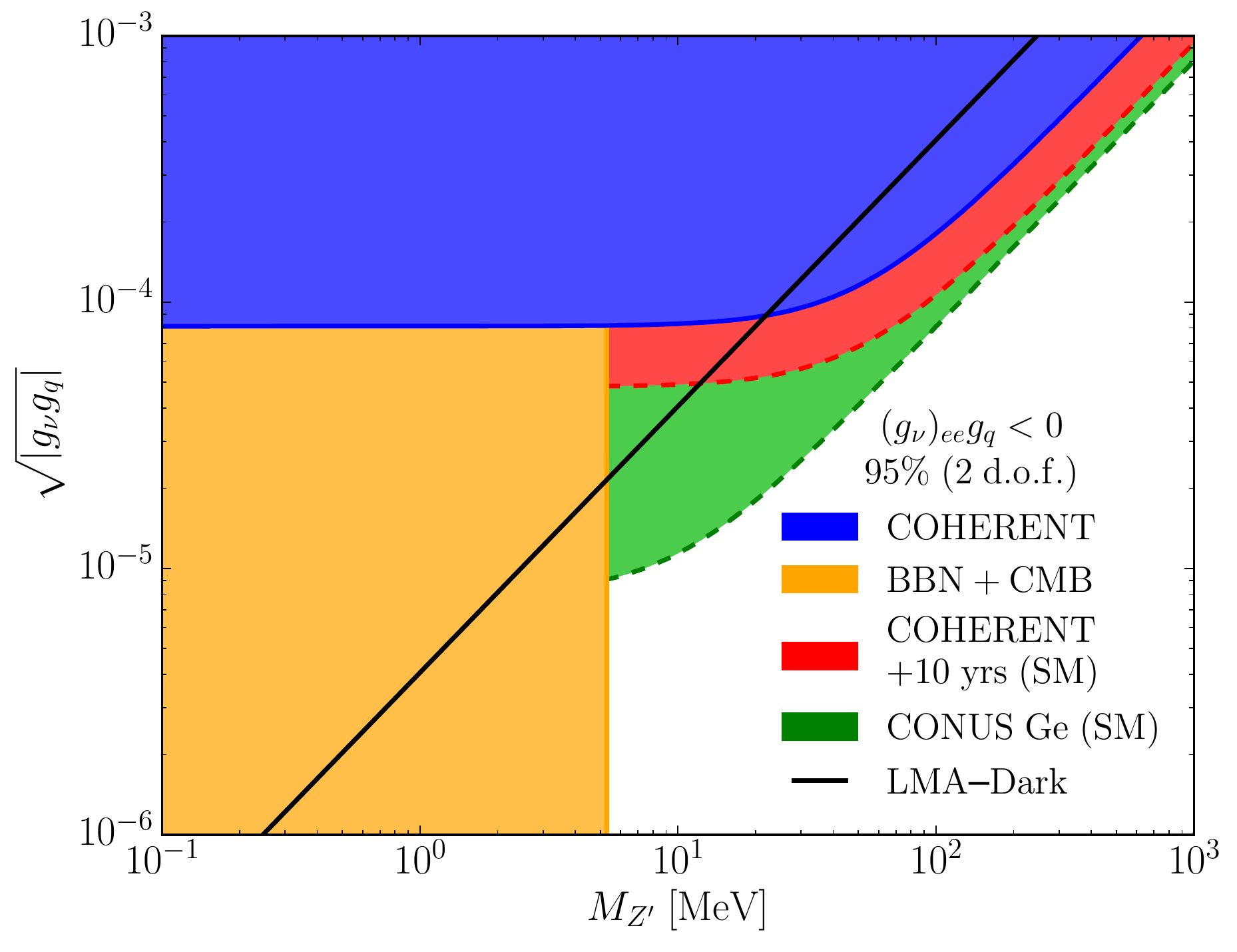}
\includegraphics[width=0.49\textwidth]{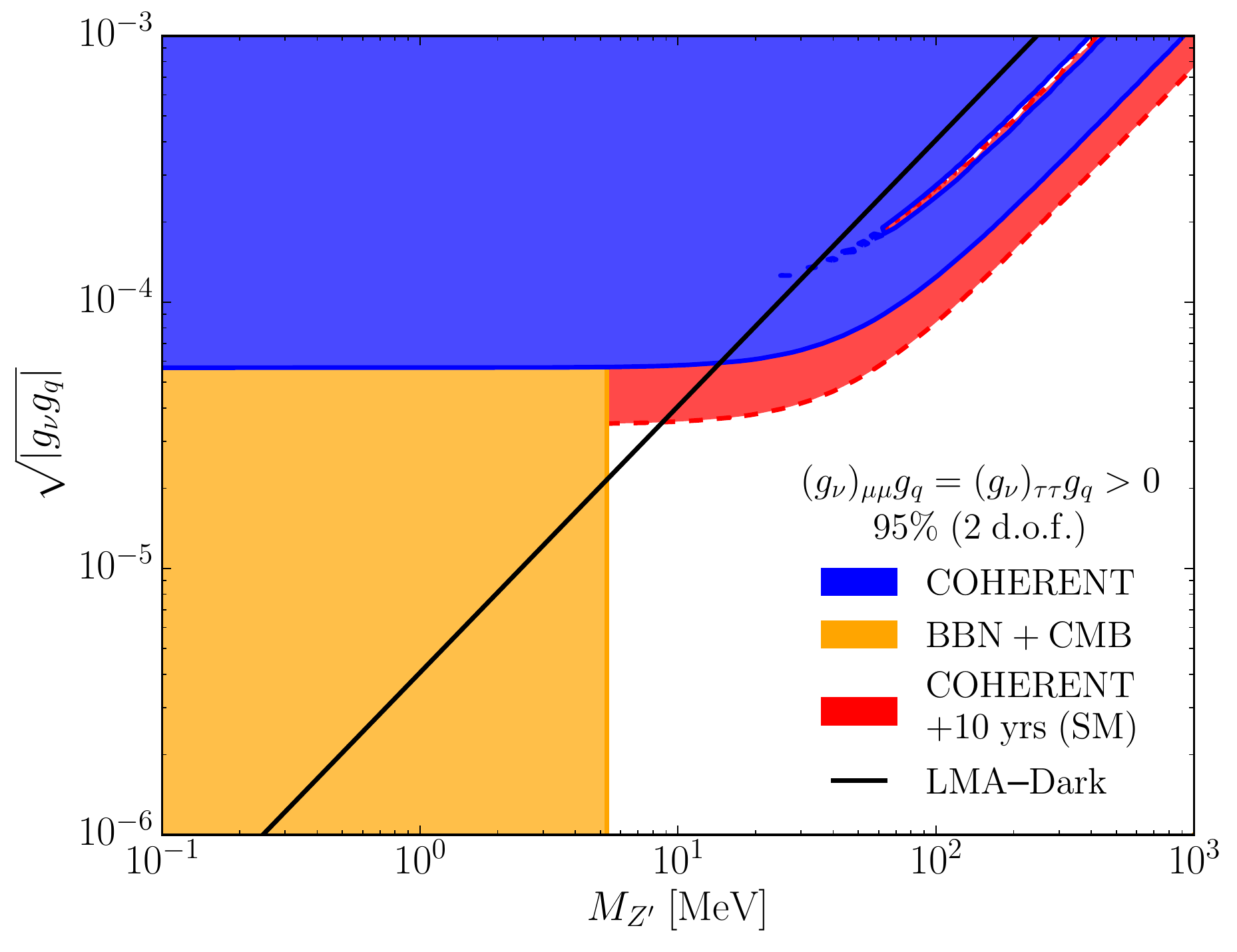}
\caption{Bounds on the product of couplings to neutrino and quark  versus the mass of mediator. In the left (right) panel, $g_\nu$ indicates $ee$ ($\mu\mu$) component.  The blue and red regions show the 95\% C.L.~with 2 d.o.f.~constraints on NSI respectively from the present COHERENT data and the forecast for 10 more years of COHERENT running  with CsI assuming no NSI. The sliver on the right panel is a result of the degeneracy in Eq.~(\ref{eq:degeneracy}). The constraint from BBN and the CMB is shown in orange \cite{Kamada:2015era}. The CONUS (see section \ref{ssec:CONUS}) constraint in green conservatively takes the Germanium detector and assumes that they will measure the SM.  CONUS cannot constrain the $\mu\mu$ or $\tau\tau$ terms. The black line in the left (right) panel correspond to the LMA--Dark solution with $x=0$ (with $x=2$).
Note that $g_\nu g_q$ is taken to be negative (positive) for the left (right) panel to give the LMA--Dark solution at $x=0$ ($x=2$).
Solid lines are current bounds, dashed lines are future bounds.}
\label{fig:g_MZ}
\end{figure}

The various constraints in the coupling--$M_{Z'}$ plane are shown in Fig.~\ref{fig:g_MZ} along with the location of the \LMAD solution. For the left figure we have only turned on the $ee$ term and have taken $(g_\nu)_{ee}g_q<0$ in agreement with the \LMAD solution at $x=0$,  for the right figure we have turned on only the $\mu\mu$ and $\tau\tau$ terms and taken $(g_\nu)_{\mu\mu}g_q=(q_\nu)_{\tau\tau}g_q>0$ in agreement with the \LMAD solution at $x=2$.
The current COHERENT constraint is shown in blue.
The thin sliver on the right figure of no sensitivity is the result of the degeneracy from Eq.~(\ref{eq:degeneracy}).
Using energy and/or timing information may be enough to rule out this sliver in the future, but whether or not this can happen is rather sensitive to the future systematics that COHERENT can reach.
Since that degeneracy only occurs for $\eps_{\alpha\alpha}^{q,V}>0$, it does not appear on the left figure of Fig.~\ref{fig:g_MZ}.
COHERENT's expected future sensitivity shown in red is for ten years of running CsI assuming 50\% uptime and that $\eps=0$.
Note that as shown in Fig.~\ref{fig:MZ_Sensitivity}, at this point COHERENT is dominated by systematics.
The orange region is the constraint from the CMB and BBN and the green region is the expected sensitivity from CONUS conservatively taken to use the Germanium detector design.
As seen from these figures while after COHERENT, still LMA-Dark with mediator in the range  $5.3~{\rm MeV}<M_{Z'}<12~{\rm MeV}$ survives,  CONUS bounds (combined with the BBN and CMB bounds)
can fully test LMA-Dark solution except for the special case $x\to 2$.

\section{Conclusions\label{CON}}
Oscillation data provides excellent constraints on new interactions in the neutrino sector parameterized as Non-Standard Interactions (NSI) for any mediator mass.
There are, however, two degeneracies from oscillation data: flavor universal contributions (parameterized as $x$ throughout this text) and the Generalized Mass Ordering Degeneracy (GMOD).
The GMOD leads to the \LMAD solution which requires interaction strength comparable to that of the weak interactions: $g^2/M_{Z'}^2\sim G_F$.
While scattering experiments can constrain both of these, they are only sensitive for mediators heavier than the characteristic energy of the experiment.
Large NSI with very light mediators $\lesssim$ 5 MeV is constrained by CMB and Big Bang Nucleosynthesis (BBN) measurements.

Thanks to COHERENT's measurement of Coherent Elastic $\nu$-Nucleus Scattering (CE$\nu$NS) with a new low-threshold CsI detector, more stringent upper bounds on the mass of the mediator for NSI can be placed than what was previously possible.
We find that the COHERENT data rule out \LMAD for $M_{Z'} > 48$ MeV at 95\% C.L.~and future measurements should improve this constraint to $\sim10$ MeV, which is not enough to close the gap with the constraints from the CMB and BBN.
However, it is possible to reach the $\sim$ MeV scale using future high statistics reactor neutrino experiments measuring CE$\nu$NS for NSI in the $ee$ sector.
With a combination of CE$\nu$NS measurements from COHERENT and reactor data along with BBN and CMB information, \LMAD in the $ee$ sector ($x\ne 2$) will be ruled for many orders of magnitude of mediator masses.
MeV scale NSI will still be viable even after reactor measurements for \LMAD NSI in the $\mu\mu$, $\tau\tau$ sector.
Notice that from model building point of view, the special case of $x=2$ is not necessarily a fine-tuned limit and can be justified by symmetries.
For example, if the new sector is electrophobic, we will expect $\epsilon_{ee}= \epsilon_{e\mu}=\epsilon_{e\tau}=0$ but still $\epsilon_{\mu \mu}, \epsilon_{\tau \tau} \ne 0$.
Until such data arrives however \LMAD will remain viable in the $\sim10$ MeV range for any $x$ and will continue to play a role in our ability to move neutrino physics into the precision era.

\acknowledgments
We thank J.J.~Cherry for useful discussions.
PBD acknowledges support from the Villum Foundation (Project No.~13164) and the Danish National Research Foundation (DNRF91 and Grant No.~1041811001).
IMS is very grateful to the Physics Department at the University of South Dakota for its support.
YF has received partial funding from the European Union\'~\!s Horizon 2020 research and innovation programme under the Marie Sklodowska-Curie grant agreement No 674896 and No 690575 for this project.
YF is also grateful to ICTP associate office for partial financial support.

\vspace{2cm}

\bibliographystyle{JHEP}

\bibliography{NSI}

\end{document}